\documentstyle{article}
\begin{document}
\newcommand{\beq}{\begin{equation}}
\newcommand{\eeq}{\end{equation}}
\newcommand{\beqn}{\begin{eqnarray}}
\newcommand{\eeqn}{\end{eqnarray}}
\newcommand{\dpf}{\displaystyle\frac}
\newcommand{\no}{\nonumber}
\newcommand{\ep}{\epsilon}
\begin{center}
{\Large Quantum entropy of two-dimensional extreme charged dilaton black
hole }
\end{center}
\vspace{1ex}
\centerline{\large Bin Wang$^{1,2}$\footnote[0]{Email: binwang@fudan.ac.cn,
Permanent address: Dept. of Physics, Shanghai Normal Univ., Shanghai
200234, P.R.China},
\ Ru-Keng Su$^{1,2}$\footnote[0]{Email: rksu@fudan.ac.cn} and P.K.N.Yu$^3$}
\begin{center}
{$^1$ China Center of Advanced Science and Technology (World Laboratory),
P.O.Box 8730, Beijing 100080, P.R.China\\
$^2$ Department of Physics, Fudan University, Shanghai 200433, P.R.China\\
$^3$ Department of Physics and Materials Science, City University of Hong
Kong,
Hong Kong}
\end{center}
\vspace{6ex}
\begin{abstract}
%\vspace{1ex}
%\begin{minipage}{130mm)
By using Hawking's treatment as well as Zaslavskii's treatment respectively
and the brick wall model, two different values of classical entropy and
quantum entropy of scalar fields in the two-dimensional extreme charged
dilaton
black hole backgrounds have been obtained. A new divergent term emerges in
the
quantum entropy under the extreme limit for Zaslavskii's treatment and its
connection
with the phase transition has been addressed.
%\end{minipage}
\end{abstract}
\vspace{6ex}
\hspace*{0mm} PACS number(s): 04.70.Dy, 04.20.Gz, 04.62.+v.
\vfill
\newpage
Two different treatments suggested by Hawking et al[1-3] and Zaslavskii [4]
respectively have been employed to calculate the classical entropy of
extremal black holes
(EBH) recently. These different treatments start with different points and
lead to very different
results. Starting with the original four-dimensional (4D)
Reissner-Nordstr$\ddot{o}$m
(RN) EBH, Hawking et al claimed that a RN EBH has zero entropy, infinite
proper distance $l$ between the horizon and any fixed
point. It cannot be formed from its nonextremal counterpart [1,5]. However,
in a grand canonical ensemble, if
first taking the boundary limit $r_+\to r_B$, where $r_+$ is the event
horizon and $r_B$ is the boundary of
the cavity, and then the extremal limit $q\to m$, Zaslavskii [4] proved
that a RN
nonextremal black hole (NEBH) can approach the extreme state as closely as
one
likes in the topological sector of nonextremal configuration. Its entropy
still obeys the Bekenstein-Hawking formula. Since these two different
treatments
approach to two different topological objects, we suggested that there are
two kinds of EBHs in
the nature. The first kind with zero entropy, zero Euler characteristic, is
the original EBH;
and the second kind with entropy $A/4$, non-zero Euler characteristic, can
be formed from its nonextremal counterpart [6].

Obviously, if these two kinds of EBHs do exist in the nature, their
different
effects will not only be demonstrated on the classical level, but also be
exhibited in
the quantum study. Many calculations, including the brick wall model
[7-10],
Pauli-Villars regular theory [11] and etc., have been used to deal with the
entropy
of quantum scalar fields on black hole backgrounds under WKB approximation
or one-loop
approximation respectively. It is believed that fields propagating in the
region just
outside the horizon give the main quantum fluctuational contribution to the
black hole
entanglement entropy [7-9,12-14]. Since these two kinds of EBHs'
backgrounds are very
different, one can predict that the quantum corrections of the scalar field
in different kinds of EBHs
will have considerable differences. This paper envolves from an attempt to
study
the quantum fluctuational corrections of these two EBHs.

To illustrate our results transparently, for simplicity, we investigate
two-dimensional
(2D) charged dilaton black hole(CDBH), which is an asymptotically flat
solution for the
heterotic string [15,16]. The action has the form
\beq                  %eq(1)
I=-\int_M
\sqrt{g}e^{-2\phi}[R+4(\bigtriangledown\phi)^2+\lambda^2-\dpf{1}{2}
  F^2]-2\int_{\partial M}e^{-2\phi}K
\eeq
which has a black hole solution metric
\beqn         %eq(2-4)
{\rm d}s^2=-g(r){\rm d}t^2+g^{-1}(r){\rm d}r^2\\
g(r)=1-2me^{-\lambda r}+q^2e^{-2\lambda r}      \\
e^{-2\phi}=e^{-2\phi_0}e^{\lambda r},\ A_0=\sqrt{2}qe^{-\lambda r}
\eeqn
where $m$ and $q$ are the mass and electric charge of the black hole
respectively.
The horizons are located at $r_{\pm}=(1/\lambda)\ln(m\pm \sqrt{m^2-q^2})$.

Using the finite-space formulation of black hole thermodynamics, employing
the
grand canonical emsemble and putting the black hole into a cavity as
usual[4,17],
we calculate the free energy and entropy of the CDBH. To simplify our
calculations, we
introduce a coordinate transformation
\beq    %eq(5)
r=\dpf{1}{\lambda}\ln[m+\dpf{1}{2}e^{\lambda(\rho+\rho_0^*)}+\dpf{m^2-q^2}{2
}
    e^{-\lambda(\rho+\rho_0^*)}]
\eeq
where $\rho_0^*$ is an integral constant, and rewrite Eq.(2) to a
particular gauge
\beq     %eq(6)
{\rm d}s^2=-g_{00}(\rho){\rm d}t^2+{\rm d}\rho^2
\eeq
The event horizon locate at $\rho_+=(1/\lambda)\ln\sqrt{m^2-q^2}-\rho_0^*$.
The Euclidean action
takes the form
\beq         %eq(7)
I=-\int_{\partial M}
\sqrt{\dpf{1}{g_{11}}}e^{-2\phi}(\dpf{1}{2}\dpf{\partial _1 g_{00}}
   {g_{00}}-2\partial_1 \phi)
\eeq
The dilaton charge is found to be
\beqn          %eq(8,9)
D & = & e^{-2\phi_0}(m+\dpf{1}{2}e^x+\dpf{m^2-q^2}{2}e^{-x})\\
x & = & \lambda(\rho+\rho_0^*)
\eeqn
The free energy, $F=I/{\beta}$, where $\beta$ is the proper periodicity of
Euclideanized time at a fixed value of the spatial coordinate and has the
form
$\beta=1/T_w=\sqrt{g_{00}}/T_c$. $T_c$ is the inverse periodicity of the
Euclidean time
at the horizon
\beq         %eq(10)
T_c=\dpf{\lambda\sqrt{m^2-q^2}}{2\pi(m+\sqrt{m^2-q^2})}
\eeq
Using the formula of entropy $S=-(\partial F/\partial T_w)_D$, we obtain
\beq             %eq(11)
S=\dpf{2\pi e^{-2\phi}[m+\dpf{e^x}{2}+\dpf{(m^2-q^2)e^{-x}}{2}][1+(m^2-
q^2)e^{-2x}]  \sqrt{m^2-q^2}(m+\sqrt{m^2-q^2})}{(m^2-
q^2)+m[\dpf{e^x}{2}+\dpf{(m^2-q^2)e^{ -x}}{2}]} 
\eeq
We are now in a position to extend the above calculations to EBH. We are
facing
two limits, namely, the extreme limit $q\to m$ and the boundary limit
\beq
x\to x_+=\lambda(\rho_+ +\rho_0^*)=\ln\sqrt{m^2-q^2}
\eeq

We can take the limits in different orders: (A) by first taking the extreme
limit
$q\to m$ and then the boundary limit. This corresponds to the treatment of
Hawking et al.
by starting with an original first kind of EBHs. (B) by first taking the
boundary
limit $x\to x_+$ and then the extreme limit $q\to m$. This corresponds to
the treatment of Zaslavskii
and obtains the second kind of EBHs. To do our limits procedures
mathematically, we may take $x=x_+ +\epsilon,
\epsilon\to 0^+$ and $m=q+\eta, \eta\to 0^+$, where $\epsilon$ and $\eta$
are
infinitesimal quantities with different orders of magnitude, and substitute
them into
Eq(11). It can easily be shown that in treatment (A)
\beq
S_{CL}(A)=0
\eeq
which is in consistent with the result in [18]. However in treatment (B)
   \beq          %eq(13)
   S_{CL}(A)=4\pi me^{-2\phi_0}
   \eeq
   which is just the Bekenstein-Hawking entropy. These results support that
there are two kinds of 2D charged
   dilaton EBHs [6] which have completely different values of classical
entropy.

   Extending our calculations to quantum cases, we suppose that the CDBH is
enveloped by
   a scalar field, and the whole system, the hole and the scalar field, are
filling
   into a cavity. The wave equation of the scalar field is
     \beq           %eq(15)
     \dpf{1}{\sqrt{-g}}\partial _\mu(\sqrt{-g}g^{\mu\nu}\partial
_\nu\phi)-m^2\phi=0
     \eeq
     Substituting the metric Eq.(2) into Eq.(15), we find
     \beq          %eq(16)
     E^2(1-2me^{-\lambda r}+q^2e^{-2\lambda
r})^{-1}f+\dpf{\partial}{\partial r}
      [(1-2me^{-\lambda r}+q^2e^{-2\lambda r})\dpf{\partial f}{\partial
r}]-m^2f=0
      \eeq
      Introducing the brick wall boundary condition[7]
      \beqn
      \phi(x)=0\  {\rm at}\ r=r_+ +\ep\no     \\
      \phi(x)=0\  {\rm at}\ r=L  \no   \\  \no
      \eeqn
      and calculating the wave number $K(r,E)$ and the free energy $F$, we
get
\beq                    %eq(17)
K^2(r,E)=(1-2me^{-\lambda r}+q^2e^{-2\lambda r})^{-1}[(1-2me^{-\lambda
r}+q^2e^{-2\lambda r})^{-1}E^2-m^2]
\eeq
\beq             %eq(18)
F_{QM}=-\dpf{\pi}{6\beta^2
\lambda}[\dpf{1}{2}\ln(R^2-2mR+q^2)+\dpf{m}{2\sqrt{m^2-q^2}}
        \ln\dpf{R-m-\sqrt{m^2-q^2}}{R-m+\sqrt{m^2-q^2}}]
\eeq
where $R=e^{\lambda(r_+ +\ep)}$, and $\ep\rightarrow 0$ is the coordinate
cutoff
parameter.
To extend the above discussion to EBH, we are facing two limits
$\epsilon\to 0$ and $q\to m$ again.
For treatment (A), taking the extreme limit first and $\epsilon\to 0$
(i.e.$r\to r_+$)
afterwards, we obtain
\beq
F_{QM}(A)=-\dpf{\pi}{6\beta^2\lambda}(\dpf{m}{m\lambda\epsilon}+\ln\dpf{1}{m
\lambda\epsilon})
\eeq
For treatment (B), adopting $\epsilon\to 0$ (i.e. $r\to r_+$) at first and
the extreme limit afterwards, we get
\beq
F_{QM}(B)=-\dpf{\pi}{6\beta^2\lambda}(\ln\dpf{1}{m\lambda\epsilon}+\dpf{m}{2
\sqrt{m^2-q^2}}
          \ln\dpf{2\sqrt{m^2-q^2}}{\lambda\epsilon(m+\sqrt{m^2-q^2})})
\eeq
For the sake of following discussion, we leave the term $\sqrt{m^2-q^2}$ in
the last term of
Eq(22) for the moment.

Using the formula $S=\beta^2(\partial F/\partial\beta)$, the entropy in two
treatments can be obtained as
\beqn
S_{QM}(A) & = &
\dpf{\pi}{3\beta\lambda}(\dpf{m}{m\lambda\epsilon}+\ln\dpf{1}{m\lambda\epsilon}) \\
S_{QM}(B) & = & \dpf{\pi}{3\beta\lambda}(\ln\dpf{1}{m\lambda\epsilon}+
               \dpf{m}{2\sqrt{m^2-q^2}}\ln\dpf{2\sqrt{m^2-
q^2}}{\lambda\epsilon(m+\sqrt{m^2 -q^2})})
\eeqn
Obviously, different treatments again lead to different entropy of the
scalar field in the black hole background. These results support that
there are two kinds of 2D charged dilaton EBHs from the quantum point of
view.

It is worth noting , in addition to the usual ultraviolet divergence
$\epsilon\to 0$
, which has been found for both 4D and 2D NEBH cases [7-14] and was
suggested capable of being overcome by different renormalization
methods [9,19], a new divergent term for Zaslavskii treatment emerges in
Eq(22).
The last term of the right-hand-side of Eq(22) diverges by taking the limit
$m\to q$. This term does not appear in Eq(21) for the treatment
of Hawking et al. To explain these results, we go over the arguments of
Hawking et al. and that of Zaslavskii again.
Due to topological differences, Hawking et al. claimed that the first kind
of EBH and
its NEBH counterpart are completely different objects. They cannot  be
transformed from each other.
This kind of EBH is the original EBH and can only be prepared at the
beginning of the universe [1].
Since we put $q=m$ first and then calculate its entropy by using the usual
thermodynamical approach [17],
natually its quantum entropy includes ultraviolet divergence only. But for
the second kind of EBH suggested by Zaslavskii, it can be
transformed as closely as one likes from its nonextremal counterpart by
taking the limit $m\to q$.
Keeping in mind that the entropy of scalar field in the EBH background was
obtained through WKB approximation, the
divergent quantum entropy due to taking $q\to m$ reflects in fact the 
divergent quantum fluctuations of the entanglement entropy
of the whole system including the EBH and the scalar field. According to
the statistical
physics, the infinite fluctuation breaks down the rigorous meanings of
thermodynamical quantities and is
just the characteristic of the point of phase transition. This conclusion
is in good agreement with previous studies about the phase transition of
black hole
[20-23]. With the aid of Landau-Lifshitz order-disorder phase transition
theory, in a series of 
previous papers [20-23], it has been found a phase transition will happen
when NEBH approaches to its
EBH counterpart. The new divergent term of the quantum fluctuation of
entropy when $q\to m$ in Eq(22)
supports the viewpoint that a phase transition will happen when a NEBH
approaches to the second kind of EBH.

In summary, we have extended two different treatments from 4D studies to 2D
CDBH and studied its
classical entropy as well as quantum entropy. In addition to the zero
classical entropy obtained by using Hawking
treatment [18], a nonzero classical result has been derived for the second
kind of EBH by Zaslavskii treatment.
We have also calculated the quantum entropy of scalar field in the EBH
background, and proved that  these two
different treatments again lead to two different results. Our results
support that there are two kinds of 2D charged dilaton EBHs [6].
We have found a new fascinating divergent term in the quantum entropy due
to
$q\to m$ in the Zaslavskii's treatment, which corresponds to a phase
transition and is in agreement with previous classical studies.

\vspace{1ex}
\hspace{0mm} This work was supported in part by NNSF of China.

\end{document}